\documentclass[pdflatex,iicol,sn-mathphys-num]{sn-jnl}

\usepackage{graphicx}%
\usepackage{multirow}%
\usepackage{amsmath,amssymb,amsfonts}%
\usepackage{amsthm}%
\usepackage{mathrsfs}%
\usepackage[title]{appendix}%
\usepackage{xcolor}%
\usepackage{textcomp}%
\usepackage{manyfoot}%
\usepackage{booktabs}%
\usepackage{algorithm}%
\usepackage{algorithmicx}%
\usepackage{algpseudocode}%
\usepackage{listings}%

\usepackage{aas_macros}

\newcommand{\un}[1]{\mbox{ \rmfamily #1}}
\newcommand{\unp}[1]{\mbox{\rmfamily #1}}

\newcommand{\fig}[1]{Fig.~\ref{#1}}

\newcommand{\YR}{2024\nobreakspace YR\textsubscript{4}}

\newcommand{\undeg}{\mbox{\textdegree}}
\newcommand{\unmin}{'}
\newcommand{\unsec}{''}
\newcommand{\unmas}{\mbox{ \rmfamily mas}}

\newcommand{\unmicron}{\mbox{\rmfamily{ \textmu m}}}
\newcommand{\simt}{\mathord{\sim}}

\begin{document}

\title[Astrometric follow-up of \YR]{Astrometric follow-up of near-Earth asteroid \YR\ during a Torino scale level 3 alert}


\author*[1]{\fnm{Marco} \sur{Micheli}}\email{marco.micheli@ext.esa.int}
\author[1]{\fnm{Maxime} \sur{Devogèle}}

\author[2]{\fnm{Larry} \sur{Denneau}}

\author[3]{\fnm{Eileen V.} \sur{Ryan}}
\author[3]{\fnm{William H.} \sur{Ryan}}
\author[4]{\fnm{Petr} \sur{Pravec}}
\author[4]{\fnm{Kamil} \sur{Hornoch}}
\author[4]{\fnm{Hana} \sur{Ku\v{c}\'akov\'a}}
\author[4]{\fnm{Petr} \sur{Fatka}}
\author[5]{\fnm{Melissa J.} \sur{Brucker}}
\author[5]{\fnm{Cassandra} \sur{Lejoly}}
\author[6]{\fnm{Nicholas} \sur{Moskovitz}}
\author[7,8]{\fnm{Mikael} \sur{Granvik}}
\author[7]{\fnm{Zuri} \sur{Gray}}
\author[9,7]{\fnm{Grigori} \sur{Fedorets}}
\author[10]{\fnm{Anlaug Amanda} \sur{Djupvik}}
\author[5]{\fnm{Carson} \sur{Fuls}}
\author[5]{\fnm{David} \sur{Rankin}}
\author[5]{\fnm{Kacper} \sur{Wierzchoś}}
\author[11]{\fnm{Bill} \sur{Gray}}
\author[12]{\fnm{Tim} \sur{Lister}}
\author[2]{\fnm{Richard J.} \sur{Wainscoat}}
\author[13]{\fnm{Robert} \sur{Weryk}}
\author[14]{\fnm{Olivier R.} \sur{Hainaut}}
\author[15]{\fnm{Federica} \sur{Spoto}}
\author[15]{\fnm{Peter} \sur{Veres}}

\author[16]{\fnm{Andrew S.} \sur{Rivkin}}
\author[17]{\fnm{Bryan J.} \sur{Holler}}
\author[18]{\fnm{Artem Y.} \sur{Burdanov}}
\author[18]{\fnm{Julien} \sur{de Wit}}
\author[19]{\fnm{Davide} \sur{Farnocchia}}

\author[1]{\fnm{Regina} \sur{Rudawska}}
\author[1]{\fnm{Eduardo} \sur{Alonso Peleato}}
\author[1]{\fnm{Francisco} \sur{Ocaña}}
\author[2]{\fnm{John} \sur{Tonry}}
\author[20]{\fnm{Jeroen} \sur{Audenaert}}

\author[1]{\fnm{Laura} \sur{Faggioli}}
\author[1]{\fnm{Francesco} \sur{Gianotto}}
\author[1]{\fnm{Marco} \sur{Fenucci}}
\author[1]{\fnm{Luca} \sur{Conversi}}
\author[1]{\fnm{Richard} \sur{Moissl}}

\affil*[1]{%
  \orgdiv{ESA NEO Coordination Centre}, 
  \orgname{Planetary Defence Office, European Space Agency}, 
  \orgaddress{%
    \street{Largo Galileo Galilei, 1}, 
    \postcode{00044}, 
    \city{Frascati}, 
    \state{RM},
    \country{Italy}%
  }%
}

\affil[2]{%
  \orgdiv{Institute for Astronomy}, 
  \orgname{University of Hawaii}, 
  \orgaddress{%
    \street{2680 Woodlawn Dr}, 
    \city{Honolulu}, 
    \state{HI}, 
    \postcode{96822}, 
    \country{United States}%
  }%
}

\affil[3]{%
  \orgdiv{Magdalena Ridge Observatory}, 
  \orgname{New Mexico Tech}, 
  \orgaddress{%
    \street{801 Leroy Pl}, 
    \city{Socorro}, 
    \state{NM}, 
    \postcode{87801}, 
    \country{United States}%
  }%
}

\affil[4]{%
  \orgname{Astronomical Institute of the Czech Academy of Sciences}, 
  \orgaddress{%
    \street{Fričova 298}, 
    \postcode{251 65}, 
    \city{Ondřejov}, 
    \country{Czech Republic}%
  }%
}

\affil[5]{%
  \orgdiv{Lunar and Planetary Laboratory}, 
  \orgname{University of Arizona}, 
  \orgaddress{%
    \street{1629 E University Blvd}, 
    \city{Tucson}, 
    \state{AZ}, 
    \postcode{85721}, 
    \country{United States}%
  }%
}

\affil[6]{%
  \orgname{Lowell Observatory}, 
  \orgaddress{%
    \street{1400 W Mars Hill Rd}, 
    \city{Flagstaff}, 
    \state{AZ}, 
    \postcode{86001}, 
    \country{United States}%
  }%
}

\affil[7]{%
  \orgdiv{Department of Physics}, 
  \orgname{University of Helsinki}, 
  \orgaddress{%
    \street{P.O. Box 64}, 
    \postcode{00014}, 
    \city{Helsinki}, 
    \country{Finland}%
  }%
}

\affil[8]{%
  \orgdiv{Asteroid Engineering Laboratory}, 
  \orgname{Lule\aa{} University of Technology}, 
  \orgaddress{%
    \street{Bengt Hultqvists v\"ag 1}, 
    \postcode{981 92}, 
    \city{Kiruna}, 
    \country{Sweden}%
  }%
}

\affil[9]{%
  \orgdiv{Finnish Centre for Astronomy with ESO}, 
  \orgname{University of Turku}, 
  \orgaddress{%
    \street{Vesilinnantie 5}, 
    \postcode{20014}, 
    \city{Turku}, 
    \country{Finland}%
  }%
}

\affil[10]{%
  \orgname{Nordic Optical Telescope}, 
  \orgaddress{%
    \street{Rbla. José Ana Fernández Pérez, 7}, 
    \postcode{38711}, 
    \city{Breña Baja}, 
    \country{Spain}%
  }%
}

\affil[11]{%
  \orgname{Project Pluto}, 
  \orgaddress{%
    \street{168 Ridge Rd}, 
    \city{Bowdoinham}, 
    \state{ME}, 
    \postcode{04008}, 
    \country{United States}%
  }%
}

\affil[12]{%
  \orgname{Las Cumbres Observatory}, 
  \orgaddress{%
    \street{6740 Cortona Dr \#102}, 
    \city{Goleta}, 
    \state{CA}, 
    \postcode{93117}, 
    \country{United States}%
  }%
}

\affil[13]{%
  \orgdiv{Physics and Astronomy}, 
  \orgname{The University of Western Ontario}, 
  \orgaddress{%
    \street{1151 Richmond St}, 
    \city{London}, 
    \state{ON},
    \postcode{N6A 3K7},
    \country{Canada}%
  }%
}

\affil[14]{%
  \orgname{European Southern Observatory},
  \orgaddress{%
    \street{Karl-Schwarzschild-Straße 2}, 
    \postcode{85748}, 
    \city{Garching bei München}, 
    \country{Germany}%
  }%
}

\affil[15]{%
  \orgdiv{Minor Planet Center}, 
  \orgname{Smithsonian Astrophysical Observatory}, 
  \orgaddress{%
    \street{60 Garden S}, 
    \city{Cambridge}, 
    \state{MA}, 
    \postcode{02138}, 
    \country{United States}%
  }%
}

\affil[16]{%
  \orgdiv{Applied Physics Laboratory}, 
  \orgname{Johns Hopkins University}, 
  \orgaddress{%
    \street{11100 Johns Hopkins Rd}, 
    \city{Laurel}, 
    \state{MD}, 
    \postcode{20723}, 
    \country{United States}%
  }%
}

\affil[17]{%
  \orgname{Space Telescope Science Institute}, 
  \orgaddress{%
    \street{3700 San Martin Dr}, 
    \city{Baltimore}, 
    \state{MD}, 
    \postcode{21210}, 
    \country{United States}%
  }%
}

\affil[18]{%
  \orgdiv{Department of Earth, Atmospheric and Planetary Sciences}, 
  \orgname{Massachusetts Institute of Technology}, 
  \orgaddress{%
    \street{77 Massachusetts Ave}, 
    \city{Cambridge}, 
    \state{MA}, 
    \postcode{02139}, 
    \country{United States}%
  }%
}

\affil[19]{%
  \orgdiv{Jet Propulsion Laboratory}, 
  \orgname{California Institute of Technology}, 
  \orgaddress{%
    \street{4800 Oak Grove Dr}, 
    \city{Pasadena}, 
    \state{CA}, 
    \postcode{91109}, 
    \country{United States}%
  }%
}

\affil[20]{%
  \orgdiv{MIT Kavli Institute for Astrophysics and Space Research}, 
  \orgname{Massachusetts Institute of Technology}, 
  \orgaddress{%
    \street{77 Massachusetts Ave}, 
    \city{Cambridge}, 
    \state{MA}, 
    \postcode{02139}, 
    \country{United States}%
  }%
}


\abstract{The discovery of \YR\ presented the planetary defense community with the most significant impact threat in almost two decades, reaching level 3 on the Torino scale. The community, now mature and well-organized, responded with a global observational effort. Astrometric measurements, forming the basis for orbital refinement and impact prediction, were a central component of this response. In this paper, we present the astrometric data collected by the international community, from the time of discovery until the object became too faint for all existing observational assets, including JWST. We also discuss the coordination role played by the International Asteroid Warning Network, and the importance of publicly available image archives to enable precovery searches.}




\maketitle

\section{Introduction}\label{Intro}

Planetary defense is a complex scientific and technical field encompassing the entire process of protecting the Earth against the threat of an asteroidal impact, from the discovery of a potentially colliding asteroid to mitigation activities in case of a confirmed impact. Despite its complexities, this process always begins with the discovery of a new asteroid or comet, and specifically with the acquisition of positional measurements of the object, the so-called astrometry. Astrometry is the direct initial input for the subsequent essential step of planetary defense, the determination of an object's orbit and its impact potential. 

The entire chain of events, from discovery to impact (or demotion of the threat), is routinely exercised at a theoretical level by the community, often through the guidance of dedicated international organizations such as the International Asteroid Warning Network (IAWN) and the Space Mission Planning Advisory Group (SMPAG) \citep{2024NatCo..15.4816K}. However, the opportunities to exercise the global preparation in real scenarios are scarce, since significant impact threats to our planet are fortunately an intrinsically rare event \citep{2021Icar..36514452H}.

For two decades, the most significant real-life event of planetary defense relevance was the discovery and subsequent impact assessment of (99942) Apophis (then known as 2004\nobreakspace MN\textsubscript{4}) which, towards the end of 2004, reached an impact probability of almost 3\%, for an impact of an object of significant size ($\simt 375 \un{m}$) only 25 years in the future \citep{2008EM&P..102..425S}. The Apophis alert only lasted a few days though, until pre-discovery observations of the object from three months before discovery were located in the Spacewatch archive, and led to the exclusion of the relatively high-probability impact scenario.

Almost exactly two decades later, a much more structured and prepared planetary defense community was faced with a new comparable scenario, after the discovery of asteroid \YR\ in the last few days of 2024. This paper will briefly discuss the astrometric coverage obtained for this object, from the time of discovery to when it became too faint for all existing observational assets. The discussion will focus on the efforts that led to the removal of \YR\ as a possible impact threat for Earth, but the data acquired on the target still have relevance at the time of this writing, due to its still non-zero impact probability with the Moon in 2032, with potential consequences of relevance for the entire Earth population \citep{2025ApJ...990L..20W}.

Coverage of other aspects of the event, including physical characterization observations, orbit determination, impact probability estimates and political reactions to the event, will be left to other publications, including companion papers in this same journal issue \citep{Farnocchia2025,Devogele2025}.

\section{Discovery apparition}

\begin{figure}[ht]
\centering
\includegraphics[width=0.45\textwidth]{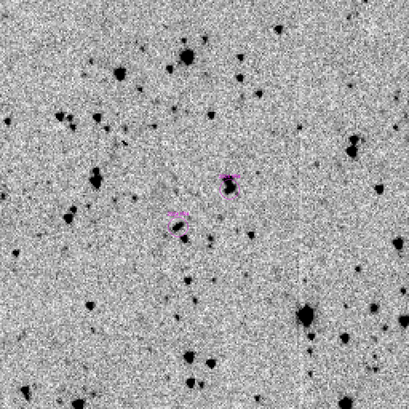}
\caption{Discovery images of \YR\ from the ATLAS station at Rio Hurtado, Chile. The two images, exposed on 2025 December 27 at 05:43 and 05:46 UTC, are stacked here without motion, displaying the asteroid as two distinct sources. The sources are slightly trailed in the motion direction, as expected for a $30\un{s}$ exposure of an object moving at $23\unsec/\unp{min}$. The pixels visible in the image are the native pixels of the camera, each $1.86\unsec$ wide, while the field of view covered by this cropped image is approximately $8\unmin \times 8\unmin$.}\label{fig:discovery}
\end{figure}

The asteroidal object now known as \YR\ was discovered by the Southern station of the ATLAS survey \citep{2018PASP..130f4505T} in Rio Hurtado, Chile (MPC code W68), on 2024 December 27, with the first observation exposed at 05:43 UTC, and a full tracklet of 4 measurements was reported to the Minor Planet Center roughly 3 hours later at 08:45 UTC. 
At the time of discovery, the object was moving at an angular speed of $23\unsec/\unp{min}$, resulting in a slightly trailed appearance in the ATLAS images (see \fig{fig:discovery}). Nevertheless, the bright $V\simt 16.5$ magnitude of the object (compared to a typical untrailed limiting magnitude of $V\simt 19.5$ for ATLAS) led to excellent astrometric measurements from the ATLAS pipeline, with reported astrometric uncertainties of the order of $\pm 0.2\unsec$ or better.

Once posted on the MPC's NEO Confirmation Page (NEOCP), follow-up observations were collected over the following 12 hours by the Catalina Sky Survey \citep{2023LPICo2851.2587C} follow-up station on Mt. Lemmon, Arizona, USA (MPC code I52), the Moriyama station in Japan (MPC code 900) and one of the iTelescope instruments in Siding Spring, Australia (MPC code Q62). Furthermore, the Catalina Sky Survey promptly found and reported prediscovery observations from their Mt. Bigelow station (MPC code 703) exposed one day earlier. This dataset was sufficient to designate the object, and it was officially announced as \YR\ on 2024 December 27 at 17:27 UTC, only 12 hours after discovery \citep{2024MPEC....Y..140D}.

Right after the discovery announcement, all impact monitoring centers (ESA's Aegis, JPL's Sentry and NEODyS) identified and published an impact possibility for 2032 December 22. Astrometric observations kept flowing over the next few days, while impact monitoring centers produced and published their updated impact monitoring assessments. For a detailed timeline of the object's rise in impact rating, we refer the reader to the companion paper in this same journal issue \citep{Farnocchia2025}.

\subsection{Global follow-up response}

In the last few days of the year, \YR\ had already risen to a Torino Scale level of 1 \citep{2000P&SS...48..297B}, attracting the attention of observers as a slightly unusual occurrence, typically happening only a few times per year. During the first weeks of 2025, while the object remained brighter than magnitude 20, multiple observing stations kept reporting astrometry, which led to the evolution of the impact probability discussed in the above-mentioned companion paper.
Dedicated professional follow-up facilities and programs also began targeting the object, obtaining and promptly reporting accurate astrometry. Most major NEO surveys also incidentally observed the object during their activities, and reported additional astrometric data.

At the same time, a similar increase of attention for this object was happening among astronomers specialized in physical-characterization observations. Lightcurve and color observations were being collected by various professional facilities, as thoroughly discussed in a second companion paper in this issue \citep{Devogele2025}. The images obtained for photometric purposes are often also excellent for astrometric use, and many observers kindly made their datasets available to extract high-precision astrometry. 

Thanks to this steady influx of observations, new high-precision astrometry of \YR, often with astrometric uncertainties of $\pm0.1\unsec$ or better, was being reported almost daily until the end of January, when the object became the first asteroid since Apophis to reach a Torino Scale level greater than 2, after crossing the threshold of $1\%$ in impact probability \citep{2000P&SS...48..297B}.

\subsubsection{Prediscovery searches and remeasurements}

Thanks to the thorough observational coverage obtained to the end of January, the ``near end'' of the observational arc at this point was extremely well-characterized. The accuracy of impact predictions was therefore dominated by the fewer earlier observations, around the time of discovery, when the object was brighter but also significantly closer, thus providing a more significant observational leverage to the orbit solution.

The community therefore initiated a careful review of these earlier observations. The original discovery tracklet was confirmed to be valid and well measured, while the prediscovery and follow-up tracklets from the Catalina Sky Survey were both remeasured with proper trail-fitting software \citep{2012PASP..124.1197V}, in order to provide a more solid estimate of the astrometric error bar associated to each detection.

Around the same time, an even earlier pair of prediscovery images, exposed on 2024 December 25, was located by the ATLAS team. 
One of the two images contained a faint and extremely trailed detection of \YR, which was nevertheless measurable. The corresponding astrometric position, despite an astrometric uncertainty of $\pm 1.6\unsec$ in the direction of motion, provided a further earlier anchoring point to the orbital solution. 
Unfortunately, the second image of that same observational set did not contain any measurable detection, since it was exposed at a time when the object was close to its lightcurve minimum, and therefore also severely diluted by trailing losses.

\subsection{IAWN involvement and further large-aperture follow-up}

\begin{figure}[ht]
\centering
\includegraphics[width=0.45\textwidth]{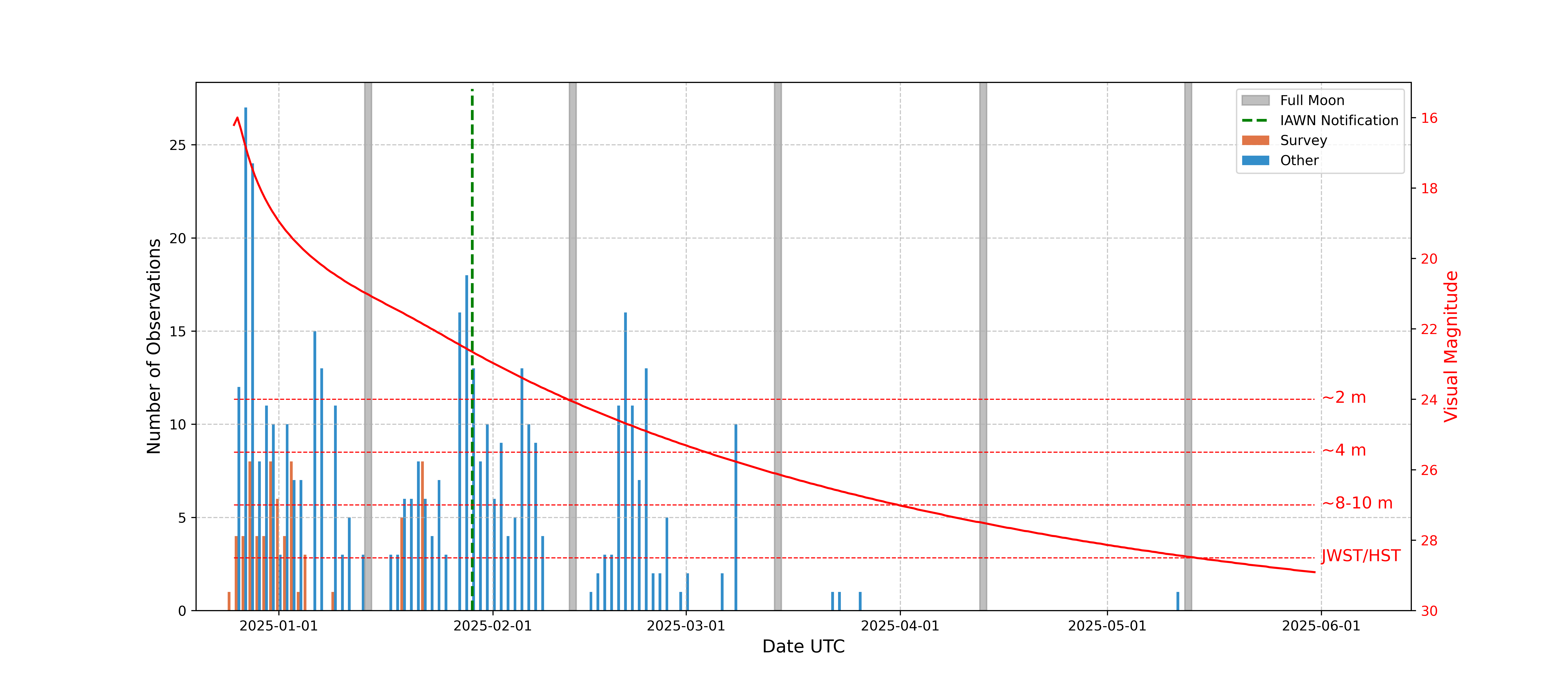}
\caption{Brightness evolution (V magnitude) of \YR\ from the time of discovery to when it became too faint even for space assets like JWST. Full moon nights, and the time of the IAWN notification, are marked on the plot. The histogram in the background indicates how many observations were reported on each UTC day. Horizontal lines indicate the approximate aperture of facilities that can meaningfully contribute astrometric measurements until that point.}\label{fig:brightness}
\end{figure}

On 2025 January 29, the IAWN released its first official potential asteroid impact notification\footnote{\url{https://iawn.net/documents/NOTIFICATIONS/2024-YR4_IAWN_Potential-Impact-Notification_20250129.pdf}}: \YR\ had now entered the spotlight as one of the most significant events in the history of planetary defense.
However, by the time of the IAWN announcement, the asteroid had already faded significantly, becoming fainter than magnitude 22 and entering a phase when only larger apertures could provide useful astrometric data (see \fig{fig:brightness}). 

Telescopic facilities with an aperture of about 2 meters, such as the Magdalena Ridge Observatory (MPC code H01), the Danish $1.54 \un{m}$ telescope (MPC code W74), the Nordic Optical Telescope (MPC code Z23) and the Faulkes Telescope North (MPC code F65) were used intensively by the corresponding observing teams, and provided crucial astrometric coverage that accompanied \YR\ during its growth up to its peak impact probability of $3\%$.\\

A full-moon break of about a week between February 9 and 15 resulted in no new observations, and by the time the object became observable again it had already faded to magnitude 24. Larger facilities, such as the Canada-France-Hawaii Telescope (CFHT, MPC code T14), now came into play to obtain valuable astrometric follow-up: due to the significant value and scarcity of larger telescopic resources, IAWN played a crucial coordination role, allowing observers to inform each other of their scheduled and executed observations, minimizing duplications and maximizing observational coverage. At the same time, the coordination ensured some redundancy between observatories remained, to avoid relying on a single station and ensuring independent verification of the astrometry.

\begin{figure}[ht]
\centering
\includegraphics[width=0.45\textwidth]{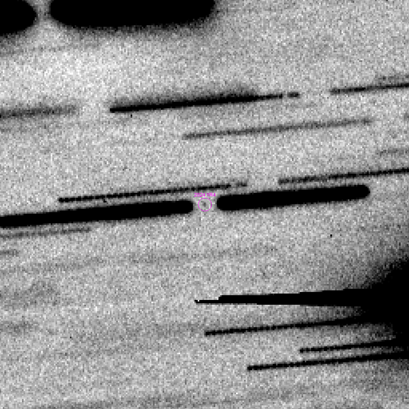}
\caption{Stack of 55 images obtained with ESO's VLT on 2025 March 23, for a total of $50\un{min}$ of exposure time. The object is the faint circled source a the center of the image. A few additional images, exposed in the same sequence, were removed from this stack because the object was transiting in the vicinity of a bright field star. This is the last ground-based optical detection of \YR\ obtained during its 2024-2025 discovery apparition.}\label{fig:VLT}
\end{figure}

These observations, collected during the second half of February, were the ones responsible for the drop in impact probability, and the subsequent full rejection of the 2032 Earth impact. Most of them were obtained with some of the largest optical telescopes on the ground, such as European Southern Observatory's Very Large Telescope (VLT, MPC code 309), the Gemini North and South telescopes (MPC codes T15 and I11), and the W. M. Keck Observatory (MPC code T16).
The last ground-based detection of \YR\ was obtained with VLT on 2025 March 23, when the object had already reached magnitude $V=26.5$ (see \fig{fig:VLT}).

\subsection{JWST extension}

By early March, with the object now fainter than magnitude 26, ground-based optical observation efforts had ceased. 

Unfortunately, the Hubble Space Telescope (HST) was unable point at the target, since it remained outside the area that can be reached with the current Two-Gyro mode of operation.

However, the James Webb Space Telescope (JWST) could point to the object, and had been shown to be uniquely suited for follow-up observations of small bodies \citep{2025Natur.638...74B} at fainter levels than what can be done optically from the ground. A successful proposal to observe the object with JWST had been submitted earlier during the apparition, when the Earth impact threat was still present \citep{2025jwst.prop.9239R}, and the corresponding observations were scheduled to be executed during the March-May timeframe.

The proposal included imaging observations in both the near-infrared (with NIRCam) and mid-infrared (with MIRI). While the latter were essential to narrow down the uncertainties on the size estimate for the object, the former were specifically designed as astrometric observations, to take advantage of JWST's exquisite resolution and sensitivity to significantly reduce the ephemeris uncertainty of \YR, had the collision with Earth still been possible. Given the existing observational arc, an orbit improvement would have required an astrometric precision better than $\simt 50 \unmas$, only reachable from a space observatory at the faintness level of \YR.\\

The complexity of an instrument like NIRCam \citep{2023PASP..135b8001R} required our team to thoroughly understand how the camera and telescope system expose, read out and attach a timetag to an imaging observation. In this work, we will not go into details of the specific technical peculiarities that have a relevance when extracting astrometry of fast-moving objects with this instrument, but we would still like to mention some of the crucial aspects to ensure the community is aware of them, and properly takes them into consideration when performing similar work on other NEOs in the future.\\

The first and foremost aspect that needs to be carefully considered when using NIRCam for astrometric measurements is the need for a sufficient number of reference stars to be detected in the field. NIRCam images processed by the JWST pipeline already include a WCS solution, but the solution has been proven good only to the level of $\simt 200\unmas$, and systematic biases at that level are almost always present \citep{2023A&A...670A..53M}. Since the goal of these observations, and in general of any astrometric observation done with JWST, is to take advantage of the exquisite resolution of the instrument to surpass ground-based astrometry, it is essential to improve upon these accuracies by astrometrically solving the images again using Gaia DR3 stars \citep{2023A&A...674A...1G, 2016A&A...595A...1G} that are detected in each frame.

\begin{figure}[ht]
\centering
\includegraphics[width=0.45\textwidth]{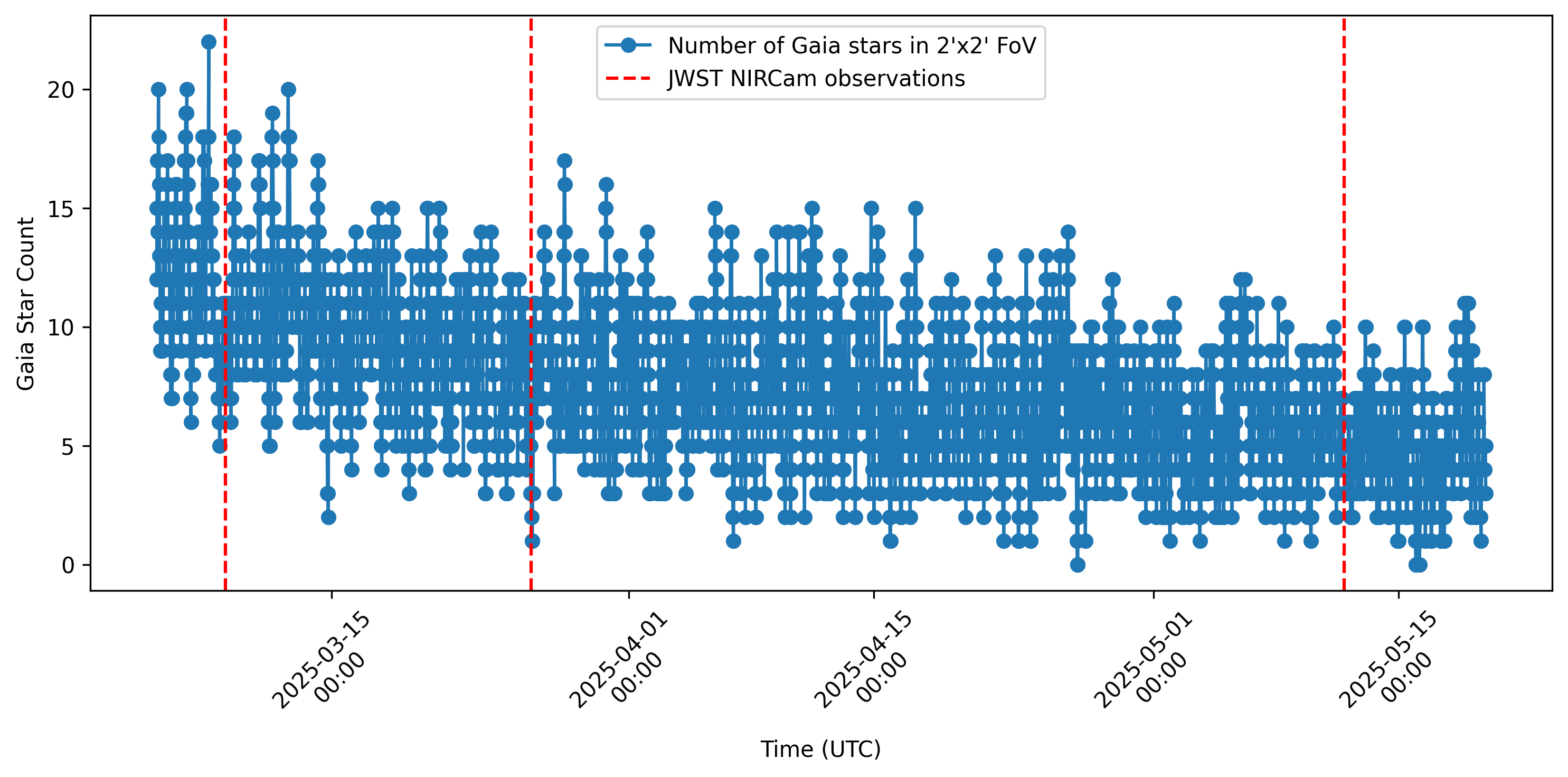}
\caption{Number of stars in the NIRCam long wavelength field of view as a function of the observing time, for the observing window used in the \YR\ campaign. The three epochs when the observations were acquired are marked by the vertical red lines.}\label{fig:GaiaStars}
\end{figure}

This task immediately presents a few issues. The first and most apparent one is the numerical density of Gaia stars in the area where the target is located. NIRCam's long wavelength channel has a field of view of just $129\unsec \times 129\unsec$ (in a single array), while the short wavelength is covered by four separate arrays, each only half that size. Consequently, it is highly likely that only a very small number of reference stars, if any, are actually within the imaged area. For this reason, during all the observed epochs of our \YR\ campaign, we had to restrict our astrometric analysis to the long wavelength channel only, despite being the least sensitive of the two for a target with the spectral profile of \YR\ (see \fig{fig:GaiaStars}).

In addition to field limitations, it is also important to notice that JWST is exposing in the near infrared, while Gaia is a visible-passband catalog: therefore, Gaia sources may not be visible in the JWST images, especially in the long wavelength channel, which covers wavelengths longer than $2.5 \unmicron$. This further reduces the number of reference sources available for the astrometric reduction.

The remaining small number of reference stars, usually between 4 and 8 for the area of the sky where \YR\ was transiting, also implied that the image could only be astrometrically solved with linear plate constants. Since the optical system of NIRCam is intrinsically more complex, and distorted at higher order, we needed to rely on the distortion corrections mapped by the JWST team, and assumed they were correct to a level sufficient for our goals. The JWST documentation \footnote{\url{https://jwst-docs.stsci.edu/jwst-calibration-status/nircam-calibration-status/nircam-imaging-calibration-status}} confirms that such corrections should be good to the level of a few mas, sufficient to not be the limiting factor in our analysis.

One last astrometric complication only became apparent once the images were made available: JWST's pipeline is designed mostly for stationary pointings, and therefore takes advantage of the fact that actual sources remain approximately in the same position on the array during the entire integration. Since IR imaging in space is affected by a large amount of variable artifacts, mostly generated by cosmic rays, the pipeline assumes that any pixel that experiences an abrupt change in flux during the exposure (and not just a linear increase with time) is being affected by an anomaly, and therefore corrects for it. However, when tracking non-sidereally, field stars enter and exit individual pixels during the exposure. This ``jump'' algorithm therefore identifies them as likely artifacts, and effectively tries to remove them from the final processed frames: the standard pipeline processing almost completely deletes the core of the streaked star PSFs from the processed frames, leaving residuals corresponding to the PSF wings. In order to avoid this issue, we had to disable the corresponding section of the JWST pipeline, with the extremely valuable help of STScI staff.\\

The next set of significant issues stems from the non-destructive up-the-ramp readout\footnote{\url{https://jwst-docs.stsci.edu/jwst-near-infrared-camera/nircam-instrumentation/nircam-detector-overview/nircam-detector-readout}} patterns\footnote{\url{https://jwst-docs.stsci.edu/jwst-near-infrared-camera/nircam-instrumentation/nircam-detector-overview/nircam-detector-readout-patterns}} used by NIRCam. For a full understanding of the process, we refer the reader to JWST's documentation\footnote{\url{https://jwst-docs.stsci.edu/understanding-exposure-times}}: for the purpose of this summary, it suffices to highlight that the overall readout of the camera effectively happens over a time window of $10.737\un{s}$, and therefore different rows of the detector have different timetags, on a timescale that cannot be neglected for high-precision astrometric purposes on an object moving at an angular speed of $\simt 10\unmas/\unp{s}$.

Furthermore, if the images are tracked non-sidereally, reference stars will also be imaged and read out at different times, depending on where they fall in the array, and therefore will be located at different positions with respect to the asteroid. This introduces a linear deformation of the astrometric solution along the direction of readout, which needs to be taken into account in the astrometric process.

In addition, each readout pattern is characterized by a different number and relative position of intervals during which the detector is not read out. In turn, these gaps in turn affect how the final image is reconstructed by the JWST pipeline, and had to be properly understood and taken into account, to avoid additional timing biases at the level of $\simt 10\un{s}$.\\

Once these peculiarities are taken into account, it is usually possible to use NIRCam detections to extract astrometry with an accuracy limited by the SNR of the detection itself.

\begin{figure}[ht]
\centering
\includegraphics[width=0.45\textwidth]{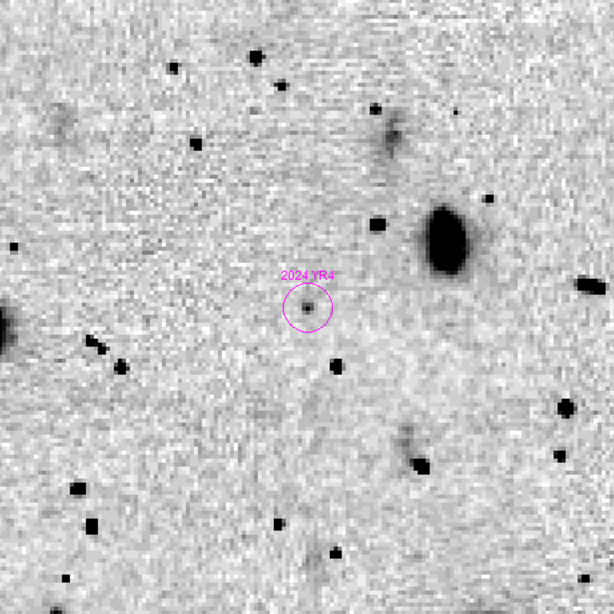}
\caption{Detection of \YR\ from the first integration of the first JWST NIRCam visit, acquired on 2025 March 8 at 21:57 UTC. The effective exposure time is approximately $107 \un{s}$. The target is circled in the frame, while the other round sources are cosmic ray hits on the array, which are not removed at processing level since the ``jump'' processing step was skipped. The brightest source in the frame is an example of a trailed field star.}\label{fig:jwst}
\end{figure}

For more information on how the observations were carried out, we refer the reader to \citep{2025RNAAS...9...70R}. For our astrometric purposes, it is only useful to know that the first visit was scheduled for 2025 March 8: the telescope successfully acquired the entire set of NIRCam observations, but the linked MIRI observation failed, resulting in the entire visit being declared as failed and scheduled to be repeated.
Nevertheless, since the NIRCam set was successfully acquired, we used it to extract a first set of astrometric measurements. The object was at the time still bright enough to be detected in each individual integration (see \fig{fig:jwst}): it was therefore possible to extract 12 individual measurements, 10 of which were unaffected by field stars or artifacts and were therefore usable. A sufficient number of reference stars, together with a good signal on the target, resulted in astrometric uncertainties ranging from $\pm10 \unmas$ in the direction perpendicular to the object's motion, to $\pm30 \unmas$ along the direction of motion.

The same observational sequence was then repeated on 2025 March 26. By that time, the object had faded one further magnitude, and therefore it was no longer easily detectable in individual integrations. The target was, however, still easily visible in the combination of all exposures, and we therefore extracted one single astrometric measurement from the overall stack. The resulting astrometric uncertainty was poorer, of the order of $\pm 40 \unmas$, due to both the additional limitations of the stacking process and the faintness of the source.

One additional astrometric visit, requested at the very end of the observability window for JWST, was finally executed on 2025 May 11. The object had faded a further 1.5 magnitudes, and was now at the limit of detectability for the long wavelenght channel of NIRCam. A $S/N \simt 2$ source was nevertheless visible in the overall stack, and when measured resulted in a position with an astrometric uncertainty of about $\pm120 \unmas$, the last measurable detection of \YR\ during the 2024-2025 apparition.

\section{Precovery opportunities}

\subsection{Suitable observational windows}

Within a few weeks of discovery, the observational dataset available for \YR\ was already sufficient to extrapolate its ephemeris to the past few decades, allowing us to explore if the position of the object could have been serendipitously covered by existing telescopic facilities at a time when it was sufficiently bright to be detectable \citep{1993MPEC....Y...05M}.

Given the ephemeris accuracy of the object, and the typical coverage of modern CCD archives, we decided to focus our search to a timespan between 1998 (the rough time of the ramp-up of NEO survey efforts) and today. We also limited our search to observational windows when the object was brighter than magnitude 24, roughly corresponding to the sensitivity limit for non-targeted observations of the largest-aperture imaging cameras currently available worldwide (DECam, MegaCam and SuprimeCam/HSC).

The orbital period of \YR\ is almost exactly 4 years, resulting in favorable  observational opportunities repeating with roughly the same cadence. The last such opportunity happened in 2020, when the object should have reached a peak brigthness of approximately 23.5. The previous opportunity, in the summer of 2016, should have been significantly better, with a peak brightness of approximately 20.5, while the earlier window in 2012 only reached magnitude 23. All other apparitions back to 1998 were fainter than magnitude 24, and therefore highly unlikely to have been detectable.

\subsection{Analyzed datasets}

We began our search using the CADC SSOIS system \citep{2012PASP..124..579G}, running a search of all possible matching fields containing the nominal ephemeris position of \YR\ and a few additional sample orbits within its ephemeris uncertainty. Despite indexing almost all professional facilities worldwide, after filtering by observable magnitude SSOIS only returned plausible hits for the 2016 apparition, and from two imaging instruments: the DECam imager on the CTIO Blanco telescope, and HSC on Subaru. DECam presented a total of 8 suitable fields covering parts of the uncertainty, exposed between 2016 August 11 and 13 UTC, while Subaru had plausible images on 2016 August 2, 5 and 9. No suitable fields were found for neither the 2020 nor the 2012 apparitions.

We carefully inspected the DECam fields by eye, overlaying the line of variation of the object's uncertainty to help guide the search for transient sources. When two or more images of the same area were available, we blinked the pair looking for a moving source, while when only one suitable image was available we downloaded a comparable frame from the same instrument and passband from other epochs in different years, and blinked them looking for sources that disappeared between the \YR\ epoch and the reference epoch. The search was performed independently by two people from our team, with different software and hardware support, and neither search turned up suitable candidates.

We performed a comparable analysis on the Subaru HSC datasets, astrometrically solving each chip of the camera array, locating the position of the line of variation on them, and inspecting the corresponding pixels. Again, no suitable sources were found.

Finally, we repeated the same analysis focusing on the segment of the uncertainty region that corresponded to the counterimage of where the object would have been if it were on a collision course with the Earth in 2032. Again, no compatible detection was found, even after stacking together all frames exposed on the same night.\\

At the same time, members of the community were searching for suitable fields in other image archives not covered by the SSOIS system.

The Catalina Sky Survey team located a set of 4 frames exposed by the $1.5 \un{m}$ Mt. Lemmon telescope on 2016 September 2, almost exactly at the time when the object should have reached the peak magnitude of the entire apparition, approximately 20.6.
The exposure footprint covered roughly 30\% of the ephemeris uncertainty at the time it was located, and included the region corresponding to the counterimage of the 2032 impact \citep{2000Icar..145...12M,2021A&A...653A.124H}. Various people from the Catalina team and outside were asked to independently inspect the frames and report possible moving candidates, but none were found. Nevertheless, the limiting magnitude of the dataset was extremely close to the expected brightness of \YR\ at the time, and therefore the non-detection could not be conclusively attributed to the absence of the object from that region\footnote{Specific requirements have been defined in order to consider a non-detection within the counterimage of an impactor as a conclusive proof of exclusion for the impact. They are presented in this document: \url{https://minorplanetcenter.net/mpcops/documentation/negative-observations/}. This image set of \YR\ did not meet the conditions to qualify as a negative observation.}.\\

Around the same time, the Pan-STARRS team \citep{2022DPS....5450401W} also thoroughly inspected their archive for all three possible apparitions, and extracted images corresponding to the location of the object in 2012, 2016, and 2020. A similar process was repeated, with in depth inspections of the frame by both members of the Pan-STARRS team and outside observers, but no suitable candidates were found.\\

The object was in the field of view of TESS \citep{2015JATIS...1a4003R}, sector 35, between 2021 February 9 and March 6: however, the brightness during this time window was only $V \simt 24.5$, while TESS limiting magnitude, even when shifting and stacking all images, only reached $V \simt 21$.

Furthermore, archival images from the amateur-led Galactic Plane eXoplanet Survey (GPX; \citep{2017JAVSO..45..127B,2021MNRAS.505.4956B}) were inspected. In principle, \YR\ would have been within reach of detection in shift-and-stack combinations of the wide-field GPX data obtained with a $279 \un{mm}$ f/2.2 telescope. However, the object’s trajectory passed approximately $2\undeg$ outside the effective field center of one of the survey pointings, placing it just beyond the area of sufficient sensitivity for precovery.

Multiple other archives were also independently searched for possible matching exposures, including all those indexed by the CATCH system \citep{2021plde.confE..33W}, all other NASA-funded NEO surveys, the Lowell Discovery Telescope, and other facilities. None led to any suitable images covering the right location at the right time.

\subsubsection{Precovery status with the current observed arc}

At the end of the 2024-2025 apparition, once all astrometric observations were available, we repeated a search for all fields still compatible with the smaller orbital uncertainty. Of all the images analyzed during the earlier part of the campaign, only one frame from the DECam imager still contains the current ephemeris uncertainty region, a 107 s, z-band image exposed on 2016 August 11 at 05:03 UTC. 

As of now, we know the position of \YR\ at the epoch of that image with a 1-sigma precision of approximately $\pm 2\unmin$, at a time when its expected visual magnitude should have been roughly $V\simt23$ \citep{Devogele2025}. The corresponding 3-sigma segment, approximately $12 \unmin$ long, straddles two adjacent chips of the camera array (chip 26 and 33), and overlaps the $\simt 1\unmin$ gap between the two chips. 
The object was moving at a speed of $1.53\unsec/\unp{min}$ at the time of the image acquisition, corresponding to a motion of $2.73\unsec$, or approximately 10 DECam pixels: the object should therefore appear as a faint streak in the image, obtained under $0.7\unsec$ seeing conditions.\\

A very careful inspection of the entire uncertainty reveals no obvious candidates. However, trailing losses, together with the sensitivity of the camera in the z-band, suggest that any detection would have been marginal at best, and more likely entirely undetectable. Furthermore, there is a significant chance that the object could be located in the gap between the two chips, and therefore undetected. Finally, the presence of a faint background source at the position corresponding to the counterimage of the Moon impact also does not allow us to confirm or discard this scenario.


\subsection{Proximity searches}

The searches above included the majority of professional telescopic archives, but might have missed some independent facilities, or non-archived professional programs, which might have been observing around the expected position of \YR\ at a time when it was bright enough to be visible in their images. This could have been particularly true for the 2016 apparition, when the object became brighter than magnitude 21, and therefore potentially detectable with smaller apertures.

In order to check for possible serendipitous detections in images of other asteroid observers, we ran all NEO astrometric observations reported to the MPC during the 2016 apparition window through a script that verified their distance from \YR\ at the time the observation was acquired, and reported matches when the reported target was within $2\undeg$ of \YR.

We then filtered the resulting hits down by comparing the magnitude of the targeted asteroid with the magnitude of \YR\ at that time: if the target was either fainter than \YR, or at most one magnitude brighter, we could have expected \YR\ to also be detectable in the frames, assuming the field of view was large enough.

The filters mentioned above only resulted in one set of hits, corresponding to objects in the Pan-STARRS images already inspected earlier. Therefore, no additional useful datasets were discovered with this approach.\\

In addition to searching for possible nearby asteroids, we also explored if any well-known and commonly observed deep-sky target was located in the vicinity of the ephemeris of \YR. No promising approaches were found, although the overall sky trajectory during the 2016 apparition led the object towards a frequently observed area of the Northern summer sky, which might have been serendipitously covered by independent observers. No news of existing image sets covering the appropriate area have emerged so far.

\section{Future observability opportunities}

At the time of this writing, the visual magnitude of \YR\ as observed from the Earth is already close to $V\simt 30$, unobservable by any optical facility.

The next two brightness maxima will be reached in 2026 and 2027, but they will both be extremely faint, with an expected peak magnitude of $V\simt 29.5$, again optically unobservable with any existing or foreseen facility.

The next chance for ground-based detections of \YR\ will only come in 2028 July, when the object is predicted to reach $V\simt 25$, well located near opposition. The ephemeris uncertainty at that time is already less than an arcminute, making the object easily detectable by any 8-10-meter-class optical facility.
The object will then become much brighter a few months later, reaching $V\simt 21$ in December. After that, it will fade over the following month, and the next observational opportunities will only happen in 2032, just before the close approach.\\

While the observational opportunities from the ground are nearly non-existent until 2028, the situation might be different for space facilities operating in the NIR: the object will be observable with the NIRCam imager on JWST in 2026, and observations during a suitable window are already foreseen \citep{2025jwst.prop.9441R}. If successful, these observations will allow a significant orbital improvement, necessary to enable early decision making for the possible lunar impact of \YR\ in 2032 \citep{2025ApJ...990L..20W}.

\section{Conclusions}

\begin{figure}[ht]
\centering
\includegraphics[width=0.45\textwidth]{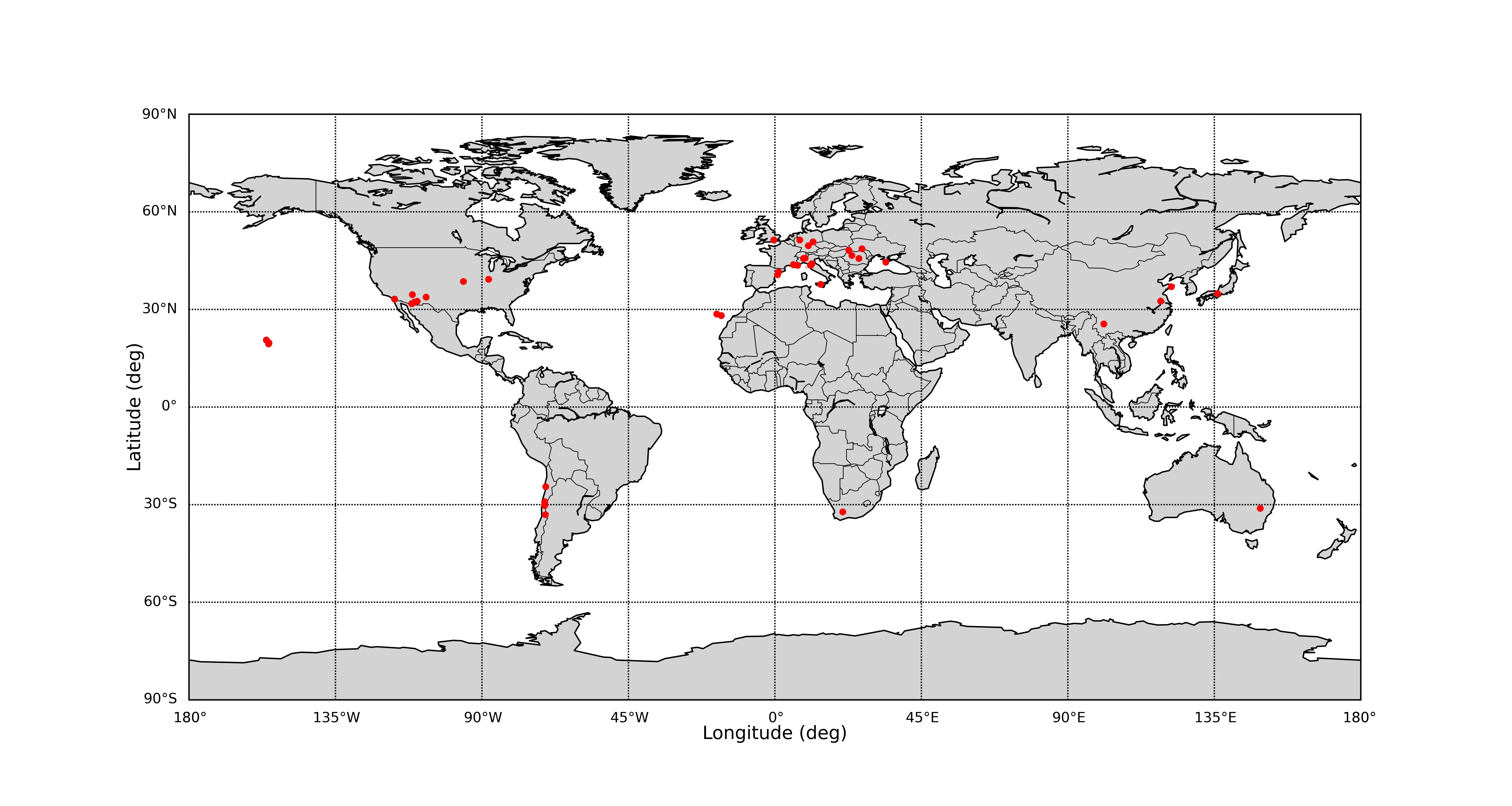}
\caption{Map of all the ground-based stations that contributed observations of \YR.}\label{fig:mappa}
\end{figure}

In this work, we have presented the astrometric components of an intensive international response to a high-profile target of planetary defense interest. In total, $504$  observations of this object, by $63$ different observing stations (see \fig{fig:mappa}), were reported over a timespan of less than six months. The outcome demonstrates that the global community of astrometric observers, under the coordination of IAWN, is capable of rapidly addressing critical observational needs, including securing access to all relevant facilities at the appropriate time.

The campaign also provided an opportunity to investigate every possible source of data and information on a critical object. This included repurposing data originally obtained for other goals as astrometric resources, and systematically examining astronomical image archives to fully assess their precovery potential. The complexity of the precovery effort discussed above nevertheless underlines the need for a more streamlined precovery mechanism in the community, which would prove valuable next time a similar object is discovered.\\

The global observational effort culminated in the downgrade and eventual removal of \YR\ as a potential impact threat to Earth, reducing the planetary defense implications of the case. Nevertheless, \YR\ remains a high-priority target due to its still non-negligible impact probability with the Moon. Such an event would not only be of exceptional scientific interest, but could also carry planetary defense implications if significant amounts of debris were ejected from the lunar surface, potentially threatening human assets in orbit around the Earth or the Moon.

Furthermore, the \YR\ alert will remain a very important example and test case of how the planetary defense community can respond to a significant impact threat. Such experience will likely become even more critical with the expected increase in the discovery rates once the Vera C. Rubin Observatory \citep{2025AJ....170...99K}, and upcoming dedicated space missions such as NEO Surveyor \citep{2023PSJ.....4..224M} and NEOMIR \citep{2024SPIE13092E..2HC}, become operational.

\bmhead{Acknowledgements}

This paper covers the contributions from observers and teams who submitted at least 4 nights of astrometric observations during the observability window of \YR. The authors nevertheless wish to acknowledge the valuable contributions of many additional observers who reported astrometric tracklets for this object to the Minor Planet Center. Each of their measurements contributed valuable constraints to the reconstruction of the trajectory of this remarkable object.

Data from the MPC's database is made freely available to the public. Funding for this data and the MPC's operations comes from a NASA PDCO grant (80NSSC22M0024), administered via a University of Maryland - SAO subaward (106075-Z6415201). The MPC's computing equipment is funded in part by the above award, and in part by funding from the Tamkin Foundation.

This work is partly based on observations collected at the European Southern Observatory under ESO programmes 113.2690.006, 113.2690.002 and 114.28HT.001. The authors are grateful to the Paranal staff for their dedication in obtaining our NEO observations for these programmes.

This work is partly based on observations made with the NASA/ESA/CSA James Webb Space Telescope. The data were obtained from the Mikulski Archive for Space Telescopes at the Space Telescope Science Institute, which is operated by the Association of Universities for Research in Astronomy, Inc., under NASA contract NAS 5-03127 for JWST. These observations are associated with program \#9239.

This work has made use of data from the European Space Agency (ESA) mission
{\it Gaia} (\url{https://www.cosmos.esa.int/gaia}), processed by the {\it Gaia}
Data Processing and Analysis Consortium (DPAC,
\url{https://www.cosmos.esa.int/web/gaia/dpac/consortium}). Funding for the DPAC
has been provided by national institutions, in particular the institutions
participating in the {\it Gaia} Multilateral Agreement.

Program GN-2025A-DD-103 based on observations obtained at the international Gemini Observatory, a program of NSF NOIRLab, which is managed by the Association of Universities for Research in Astronomy (AURA) under a cooperative agreement with the U.S. National Science Foundation on behalf of the Gemini Observatory partnership: the U.S. National Science Foundation (United States), National Research Council (Canada), Agencia Nacional de Investigaci\'{o}n y Desarrollo (Chile), Ministerio de Ciencia, Tecnolog\'{i}a e Innovaci\'{o}n (Argentina), Minist\'{e}rio da Ci\^{e}ncia, Tecnologia, Inova\c{c}\~{o}es e Comunica\c{c}\~{o}es (Brazil), and Korea Astronomy and Space Science Institute (Republic of Korea).

Based on observations made with the Nordic Optical Telescope, owned in collaboration by the University of Turku and Aarhus University, and operated jointly by Aarhus University, the University of Turku and the University of Oslo, representing Denmark, Finland and Norway, the University of Iceland and Stockholm University at the Observatorio del Roque de los Muchachos, La Palma, Spain, of the Instituto de Astrofisica de Canarias. The NOT data were obtained under program ID P68-803. The data presented here were obtained in part with ALFOSC, which is provided by the Instituto de Astrofisica de Andalucia (IAA) under a joint agreement with the University of Copenhagen and NOT.

This paper includes data collected by the TESS mission, which are publicly available from the Mikulski Archive for Space Telescopes (MAST). Funding for the TESS mission is provided by NASA's Science Mission directorate.

Funding for the data taken by the Magdalena Ridge Observatory (MRO) is supported by NASA PDCO grant 80NSSC24K0324.

The observations by P. Pravec and his team with the $1.54 \un{m}$ Danish Telescope on La Silla were supported by the {\it Praemium Academiae} grant (no. AP2401) from the Academy of Sciences of the Czech Republic.

N.M. acknowledges support from NASA YORPD grant 80NSSC21K1328, awarded to the Mission Accessible Near-Earth Object Survey (MANOS).

D.F. conducted this research at the Jet Propulsion Laboratory, California Institute of Technology, under a contract with the National Aeronautics and Space Administration (80NM0018D0004).

\bmhead{Contributions Statement}

M.M. wrote the manuscript, coordinated astrometric observations, and measured a significant part of the astrometric positions used in this work, including those from JWST.
M.D. supported M.M. during the entire observation campaign, and especially with the precovery and JWST work.
L.D. and J.T. were part of the ATLAS discovery, and the subsequent precovery searches in the ATLAS archive.
E.V.R., W.H.R., P.P., K.H., H.K., P.F., M.J.B., C.L., N.M., M.G., Z.G., G.F., A.A.D., C.F., D.R., K.W., B.G., T.L., R.J.W., R.W. and O.R.H. contributed significantly to the observational follow-up for this object.
F.S. and P.V. provided crucial support from the MPC side during the collection and submission of astrometric observations.
M.M., M.D., A.S.R, B.J.H., A.Y.B., J.d.W. and D.F. contributed significantly to the astrometric part the JWST observations.
M.M., M.D., R.R., E.A.P., F.O., J.T. and J.A. contributed to the extensive precovery searches presented in this work.
L.F., F.G., M.F., L.C. and R.M. provided support on ESA NEOCC side for all the needs that arose during the observational campaign.
M.M., A.Y.B. and M.F. prepared the figures.
All authors reviewed the manuscript.

\bibliography{sn-bibliography}

\end{document}